\documentclass[a4paper,11pt]{article}
\usepackage{pos}
\usepackage{epsfig}
\usepackage{graphicx}
\usepackage[english]{babel}
\usepackage{hyphenat}
\usepackage{amsmath}
\usepackage{mathtools}
\usepackage{mathrsfs}
\usepackage{bbm}
\usepackage{bm}
\usepackage{slashed}
\usepackage{epstopdf}
\usepackage{booktabs}
\usepackage{float}
\usepackage{placeins}
\usepackage{xfp}
\usepackage[table]{xcolor}
\definecolor{lcolor}{rgb}{0.,0.0,0.}
\definecolor{citcolor}{rgb}{0,0.,0.5}
\definecolor{darkgreen}{rgb}{0.0,0.5,0.0}
\usepackage{multirow}
\usepackage{ltablex}
\usepackage{soul}
\usepackage{makecell}
\usepackage{ulem}
\usepackage{cleveref}

\def\be{\begin{eqnarray*}}
\def\ee{\end{eqnarray*}}
\def\beq{\begin{eqnarray}}
\def\eeq{\end{eqnarray}}

\newcommand{\bea}{\beq \begin{aligned}}
\newcommand{\eea}{\end{aligned}\eeq}

\newcommand{\csq}{C^2}

\def\be{\begin{eqnarray*}}
\def\ee{\end{eqnarray*}}
\def\beq{\begin{eqnarray}}
\def\eeq{\end{eqnarray}}

\title{CGC Fit to HERA DIS and Forward Hadron Production in p+p Collisions}
\ShortTitle{CGC Fit to HERA DIS and Forward Hadron Production in p+p Collisions}

\author[a]{Piotr Korcyl}
\author*[b]{Truong My Hau Le}
\author[c,d,e]{Farid Salazar}
\author[a]{Tomasz Stebel}

\affiliation[a]{Institute of Theoretical Physics, Jagiellonian University, ul. Lojasiewicza 11, 30-348 Krak\'{o}w, Poland}
\affiliation[b]{Doctoral School of Exact and Natural Sciences, Jagiellonian University\\
prof. {\L}ojasiewicza 11, 30-348 Krak\'ow, Poland}
\affiliation[c]{Department of Physics, Temple University, Philadelphia, Pennsylvania 19122, USA}
\affiliation[d]{RIKEN-BNL Research Center, Brookhaven National Laboratory, Upton, New York 11973, USA}
\affiliation[e]{Physics Department, Brookhaven National Laboratory, Upton, New York 11973, USA}

\emailAdd{piotr.korcyl@uj.edu.pl}
\emailAdd{truongmyhau.le@doctoral.uj.edu.pl}
\emailAdd{farid.salazar@temple.edu}
\emailAdd{tomasz.stebel@uj.edu.pl}

\abstract{
Many experiments have provided possible hints of gluon saturation, including geometric scaling in deep inelastic scattering (DIS) at HERA, the suppression of forward particle yields in d+Au collisions relative to p+p collisions at RHIC, and the suppression of away-side peaks in di-hadron correlations at forward rapidities in proton-nucleus collisions at RHIC and the LHC. Previous studies within the Color Glass Condensate (CGC) framework have primarily focused on fits to HERA deep inelastic scattering data, using the resulting dipole amplitudes to predict hadronic observables in proton–nucleus and proton–proton collisions. A simultaneous fit of DIS and forward particle production data, and a systematic study of their interplay, has so far been absent. 
In the work presented here, we focus on two processes: inclusive DIS and forward single inclusive hadron production in p+p collisions, incorporating data from HERA, RHIC, and the LHC. We solve the Balitsky-Kovchegov (BK) small-x evolution equation and perform a global fit to the combined datasets. Our analysis accounts for uncertainties in fragmentation functions and provides final uncertainties for the fitted parameters. The outcomes demonstrate the potential of the CGC formalism to provide a universal description of these diverse processes at the level of the dipole amplitude. In these proceedings, after very briefly summarizing the setup and our results,
we highlight one particular aspect of the analysis, namely the complete uncertainty treatment. We validated the final uncertainties with two independent methods: the Hessian method and the Bayesian inference method, finding very good agreement.
}

\FullConference{
33rd International Workshop on Deep Inelastic Scattering and Related Subjects, 4-8 May 2026, Bologna, Italy
}

\begin{document}
\maketitle

The partonic interpretation of Quantum Chromodynamics (QCD) has been grounded by many experiments, Deep Inelastic Scattering (DIS) at HERA in particular. It is now accepted that since the number of gluons carrying a small fraction of their parent hadron's momentum considerably increases at high energies, the linear evolution equations, DGLAP \cite{Gribov:1972ri,Altarelli:1977zs,Dokshitzer:1977sg} and BFKL \cite{Lipatov:1976zz,Kuraev:1977fs,Balitsky:1978ic}, must be improved by non-linear
effects~\cite{Gribov:1984tu,Mueller:1985wy,McLerran:1993ni}. That effect generates a
new dynamical scale, the saturation scale $Q_s(x)$. Its precise determination remains difficult due to competing QCD effects that may obscure the picture \cite{Morreale:2021pnn}; hence, it is one of the major experimental goals of facilities such as the Relativistic Heavy Ion Collider and the Large Hadron
Collider, and is a central objective of the future Electron-Ion
Collider~\cite{AbdulKhalek:2021gbh}. The Color Glass Condensate
(CGC)~\cite{Gelis:2010nm} provides an effective theory for this regime of QCD.
Partonic cross-sections in the CGC are expressed as convolutions of Wilson-line
correlators with perturbatively calculable impact factors. While the initial
condition of these correlators is non-perturbative, their high-energy/rapidity
evolution is perturbatively calculable and dictated by non-linear
renormalization group equations, notably the Balitsky--Kovchegov (BK) equation and
the Jalilian-Marian--Iancu--McLerran--Weigert--Leonidov--Kovner (JIMWLK)
equation~\cite{Balitsky:1995ub,Kovchegov:1999yj,JalilianMarian:1997jx,JalilianMarian:1997gr,Kovner:2000pt,Iancu:2000hn,Iancu:2001ad,Ferreiro:2001qy}.

CGC can describe various observables in different colliding systems~\cite{Albacete:2014fwa,Morreale:2021pnn}.
In these proceedings, we summarize Ref.~\cite{Korcyl:2026nrz}, where the reduced
cross-section for DIS and the forward single
inclusive hadron production (SIHP) in proton-proton collisions were considered simultaneously. They depend on the
dipole amplitude in the fundamental and adjoint representation $R$ of the $SU(3)$ group,
\begin{align}
    \mathcal{D}_{R}(r_T, x) = \frac{1}{N_c} \left \langle \mathrm{Tr}\left[ V_R(\boldsymbol{x}) V^\dagger_R(\boldsymbol{y}) \right] \right \rangle_x\,,
\end{align}
The goal of the work was to perform
the first simultaneous CGC-based fit of the reduced DIS cross-section from HERA
and SIHP in proton-proton collisions from RHIC and LHCb. The analysis was
restricted to the leading-order (LO) impact factor for both observables, with
the running-coupling BK equation for the evolution of the dipole amplitude. 
$K$-factors were introduced to account for higher-order corrections.
NLO corrections for SIHP are expected to be larger at RHIC than at the LHC~\cite{Shi:2021hwx}; therefore, different $K$-factors for RHIC and the LHC were used. At LO in the CGC, the DIS cross-section can be obtained from the optical theorem,
\begin{equation}
\label{eq. dis}
\sigma_{T,L}^{\gamma^* p}(x_{\mathrm{Bj}},Q^2) = \frac{\sigma_0}{2}  \int  \mathrm{d}^2 \boldsymbol{r} \ 2 \mathcal{N}_F(r, x_{\mathrm{Bj}})  \int \mathrm{d} z |\psi_{T,L}^{\gamma^* \rightarrow f \bar{f}}(z,r,Q^2)|^2 \,.
\end{equation}
$\mathcal{N}_F(r, x_{\mathrm{Bj}})=1-\mathcal{D}_F(r, x_{\mathrm{Bj}})$, $\psi_{T,L}^{\gamma^* \rightarrow f \bar{f}}(z,r,Q^2)$ is the light-cone wave function for a photon with virtuality $Q^2$ fluctuating into a $f\bar{f}$ dipole of transverse size $r$, with the quark carrying longitudinal momentum fraction $z$ of the virtual photon. The reduced cross-section is given by $\sigma_{r} (y,x,Q^2) = F_2(x,Q^2) - \frac{y^2}{1+(1-y)^2} F_L(x,Q^2)\,,$
where $y=W^2/s$ is the DIS inelasticity, and $s$ is the squared center-of-mass energy of the electron-proton system and the structure functions are defined as $F_2(x,Q^2) = \frac{Q^2}{4 \pi^2\alpha_{em}}(\sigma_T^{\gamma^* p}+\sigma_L^{\gamma^* p})$ and $ F_L(x,Q^2) = \frac{Q^2}{4 \pi^2\alpha_{em}}\sigma_L^{\gamma^* p}$. The differential cross-section for SIHP in $pp$ collisions can be
written in $k_T$-factorized form as a convolution of the unintegrated gluon
distributions of the two colliding protons~\cite{Blaizot:2004wu,Blaizot:2004wv}.
In the dilute-dense approximation, the large-$x$
partons in one of the protons are treated as collinear and are described by parton
distribution functions (PDFs) $f_{i/p}$. The scattering of these partons with the low-$x$ gluon field is encoded in the Fourier transform of the dipole, $\widetilde{\mathcal{D}}_{R}(\boldsymbol{k},x) = \int \mathrm{d}^2 \boldsymbol{r}\ e^{-i \boldsymbol{k} \cdot \boldsymbol{r}} \mathcal{D}_{R}(\boldsymbol{r}, x)$.
The scattered parton $i$ fragments with the collinear fragmentation function  $D_{h/i}$ into the hadron $h$. The differential cross-section at leading order in the dilute-dense framework reads~\cite{Dumitru:2005gt,Chirilli:2012jd}:
\begin{equation}
\label{eq. sihp}
 \frac{\mathrm{d} \sigma_h}{\mathrm{d} y_h \mathrm{d}^2\boldsymbol{p}_T} = \frac{K}{(2\pi)^2} \frac{\sigma_0}{2} \int_{x_F}^1 \frac{\mathrm{d} z}{z^2} \sum_i x_1f_{i/p}(x_1,\mu^2) \widetilde{\mathcal{D}}_{F,A} \left(\frac{p_T}{z}, x_2\right) D_{h/i}(z,\mu^2)\,,
\end{equation}where $x_1 =
\frac{p_T}{z\sqrt{s}}e^{y_h} \sim 1$ and $x_2 = \frac{p_T}{z\sqrt{s}}e^{-y_h}
\ll 1$, and $i$ runs over quark flavors and the gluon. 
For a quantitative comparison between
$k_T$-factorization and the dilute-dense approximation,
see~\cite{Fujii:2026ccu}. 
The lower integration limit is given by $x_F = \frac{p_T}{\sqrt{s}} e^{y_h}$, which represents the minimum fraction of the projectile's momentum carried by the final-state hadron.
The differential yield is obtained as $\mathrm{d} N_h/ (\mathrm{d} y_h \mathrm{d}^2\boldsymbol{p}_T) = \frac{1}{\sigma_{inel}} \mathrm{d} \sigma_h /(\mathrm{d} y_h \mathrm{d}^2\boldsymbol{p}_T) $, where $\sigma_{inel}$ represents the inelastic p-p cross-section.

We parameterize the initial dipole amplitude using the $\text{MV}^\gamma$ model~\cite{McLerran:1997fk,Albacete:2010sy,Lappi:2013zma}:
\begin{equation}
    \mathcal{N}_F(r, x_0) = 1 - \exp \left[-\frac{(r^2 Q_{s0}^2)^\gamma}{4} \ln\left(\frac{1}{r \Lambda_{\text{QCD}}} +  e\right)\right]\,,
\end{equation}
where we have chosen $x_0 = 0.015$ to accommodate all data sets. $Q_{s0}^2$ and $\gamma$ are fit parameters. $\Lambda_{\text{QCD}} = 0.241$~GeV. The adjoint dipole is obtained from the fundamental one using the large-$N_c$ limit.
For the evolution, we use Balitsky's prescription for the running coupling \cite{Balitsky:2006wa} and/or include the kinematical constraint~\cite{Motyka:2009gi,Beuf:2014uia}.
We freeze the coupling in the infrared region at $\alpha_s \approx 0.76$. The fit also includes the parameters $C^2$ in the running coupling prescription \cite{Beuf:2020dxl} and the overall normalization factor $\sigma_0/2$ in Eqs.~\eqref{eq. dis} and \eqref{eq. sihp} as well as the different $K$-factors at RHIC and the LHC in the SIHP sector. The fit used the \texttt{cteq6l1} set~\cite{Pumplin_2002} for PDFs and \texttt{NNFF1.0}~\cite{Bertone:2017tyb} for FFs. We choose the factorization scale as $\mu^2 = (p_T/z)^2 + Q_s^2(\eta)$;  $Q_s^2(\eta)$ is parametrized based on the GBW model~\cite{Golec-Biernat:1998zce}. For more details on the fitting setup, see Ref.\cite{Korcyl:2026nrz}. 
We performed a global analysis of HERA~\cite{H1:2015ubc} reduced cross-sections alongside forward SIHP at RHIC (BRAHMS~\cite{BRAHMS:2004xry} and STAR~\cite{STAR:2006dgg}) and LHCb ($\sqrt{s}=5$~\cite{LHCb:2021vww} and $13$~TeV~\cite{LHCb:2021abm}). For BRAHMS data, we adopt the inelastic cross-section $\sigma_{\text{inel}} = 41$~mb~\cite{BRAHMS:2004xry} to convert our cross-section into the measured yield. We apply kinematic cuts of $Q^2 \le 45$~GeV$^2$ for DIS and $p_T \geq 1$~GeV for SIHP. To isolate the impact of the SIHP data, we compared a six-parameter global fit with a four-parameter baseline restricted to DIS data. The analysis evaluates two evolution setups: the running-coupling BK equation with (kcBK) and without (rcBK) the kinematical constraint \cite{Ducloue:2019ezk}. Both evolution setups describe the DIS and SIHP data well; see Table~\ref{tab:fit_results_params}. The kcBK-based fit prefers values of $\csq$ closer to $e^{-2\gamma_E}$, which is the natural choice in the NLO calculation \cite{Balitsky:2006wa}, whereas larger values of $\csq$ are required in the rcBK-based fit to slow down the evolution. A similar pattern was observed in NLO fits to HERA data~\cite{Beuf:2020dxl,Casuga:2025etc}.

\begin{figure}[]
	\centering
    \includegraphics[width=\textwidth]{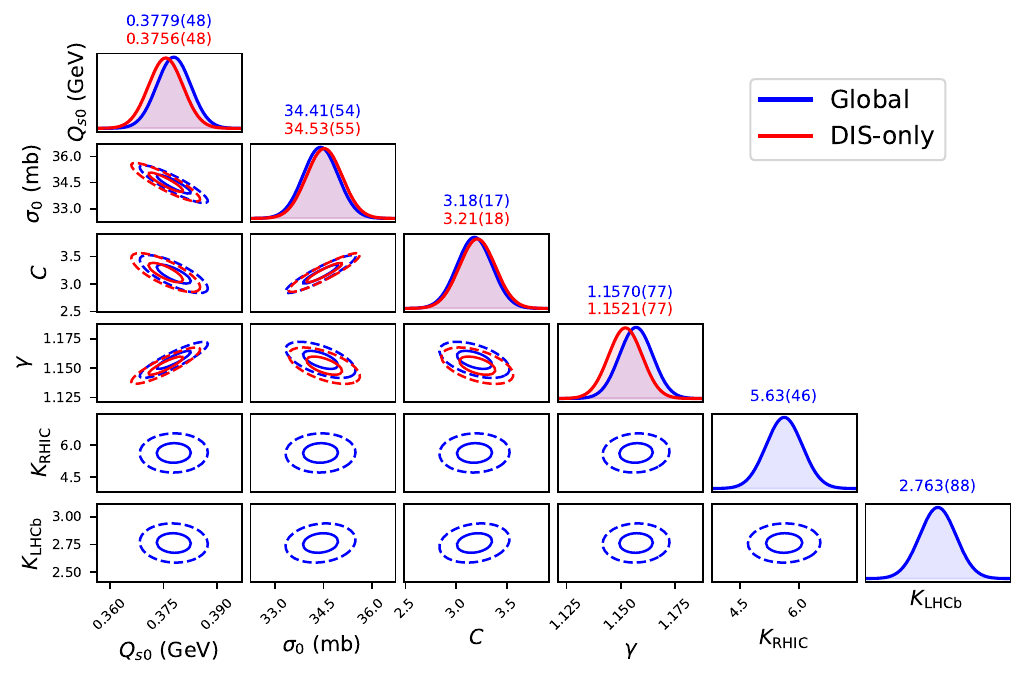}
    \caption{Parameter correlations extracted from the Hessian matrix from the global and DIS-only fits using rcBK evolution. The off-diagonal panels display $1\sigma$ (solid) and
	$2\sigma$ (dashed) confidence ellipses.}
    \label{fig:correlation}
\end{figure}

\begin{figure}[]
	\centering
    \includegraphics[width=1.0\textwidth]{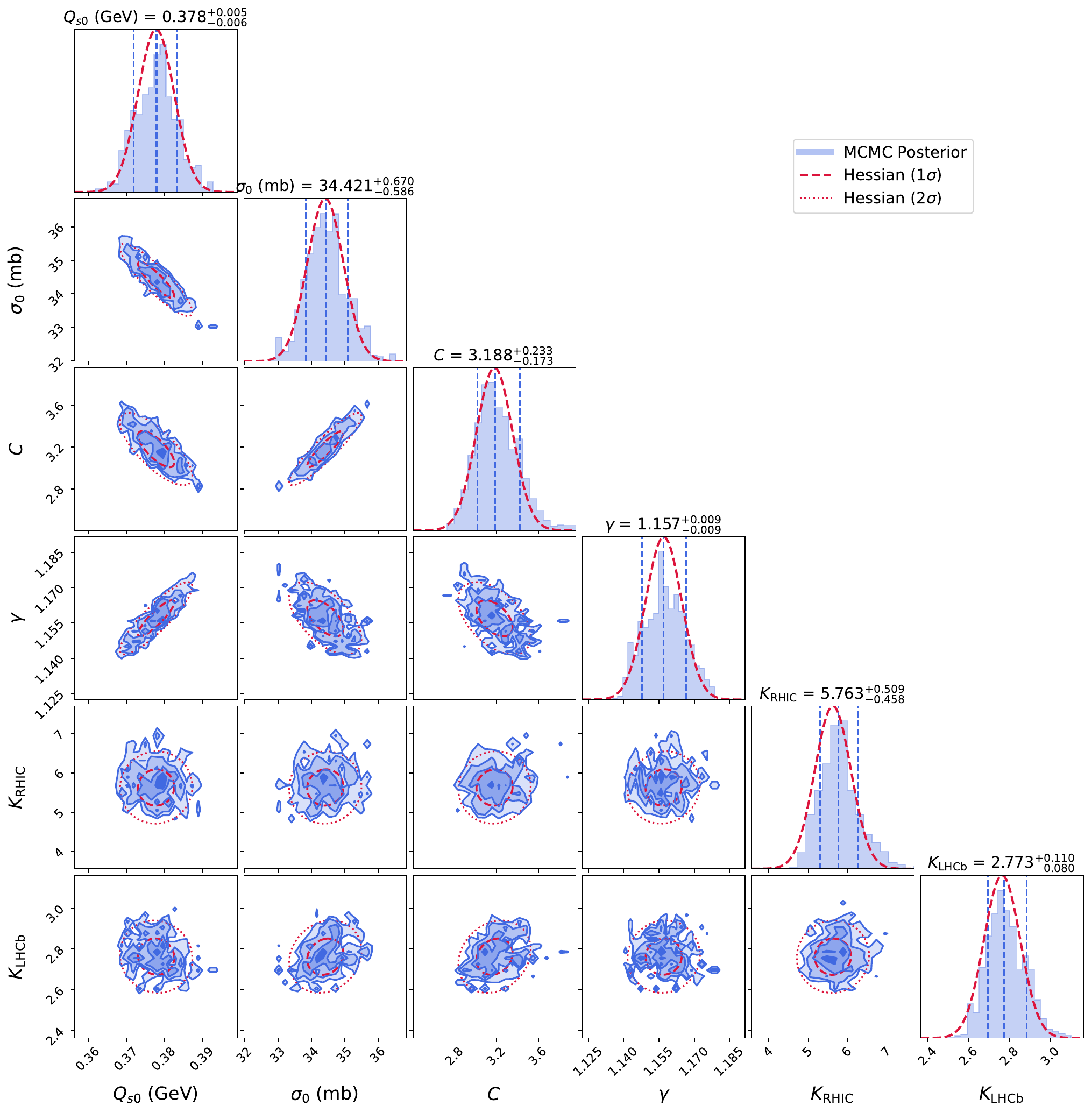}
    \caption{The comparison of posterior distributions for fitted parameters obtained using the two methods discussed in the text: Bayesian inference based on Markov Chain Monte Carlo and the Hessian method. The very good agreement between the distributions confirms the Gaussian shape of these distributions and the applicability of the Hessian method.}
    \label{fig:mccheck}
\end{figure}

The uncertainties include the propagated statistical uncertainties of the FFs, averaged over the \texttt{NNFF1.0} replica ensemble, and the estimate of systematic theoretical errors obtained through three-point scale variations, $\mu^2 \in [\mu^2/2, 2\mu^2]$. These uncertainties were included in the fitting $\chi^2$ function and were propagated to the final results of the fit parameters using two independent methods. The first one uses the  Hessian error estimation~\cite{Pumplin:2000vx,Pumplin:2001ct}. The results for the rcBK setup, are shown in Fig.~\ref{fig:correlation}. The second method is an independent Monte Carlo Bayesian analysis. The two
methods do not have free parameters and are based on different principles: the Hessian matrix is computed exactly at the minimum using automatic
differentiation \cite{Cougoulic:2024jnd}, while the Bayesian analysis samples the posterior distribution directly from the likelihood. We plot together the distributions for the fitted parameters and their correlations in Fig. \ref{fig:mccheck}. The
agreement between the two, both for the individual parameter distributions and for the correlation ellipses, confirms
the Gaussian behavior of the fitted parameters. 
This demonstrates that the determination relying only on
the Hessian matrix is adequate for global
analyses of this kind.

\begin{table}[H]
\centering
\begin{tabular}{l c c c c c c c}
\hline\hline
 & $Q_{s0}^2$ (GeV$^2$) & $\gamma$ & $C^2$ & $\sigma_0/2$ (mb) & $K$-RHIC & $K$-LHCb & $\chi^2_{\text{tot}}/\textrm{dof}$ \\
\hline
\textbf{rcBK} &  0.1411(36) & 1.1521(77)& 10.3(1.1) & 17.26(28) & --- & --- & 0.975 \\
\hline
\textbf{kcBK} & 0.1084(27) &1.1135(73)& 1.276(68) & 19.69(31) & --- & --- & 1.132 \\
\hline
\textbf{rcBK} & 0.1428(36) & 1.1570(77) & 10.1(1.1) & 17.21(27) & 5.63(46) & 2.763(88)  & \textbf{0.850} \\
\hline
\textbf{kcBK} & 0.1107(27) & 1.1210(72) & 1.254(67) & 19.57(31) & 6.03(48) &3.049(95) & 0.996 \\
\hline\hline
\end{tabular}%

\caption{Values of the fit parameters: rows 1–2 are the DIS-only baseline; the rest are the global fit.}
\label{tab:fit_results_params}
\end{table}

\begin{acknowledgments}
We thank F. Cougoulic, H. Mantysaari, and W. Zhao for valuable discussions related to this work. F.S. also acknowledges valuable discussions with Zhong-Bo Kang and Amanda Wei that motivated this project.  Numerical calculations were performed on the LUMI supercomputer under the time allocations: project\_465002091 (Calculating predictions for EIC physics) and project\_465002778 (Improving the precision of predictions for EIC physics). We gratefully acknowledge the Polish high-performance computing infrastructure PLGrid (HPC Center: ACK Cyfronet AGH) for providing computer facilities and support within the computational grant no. PLG/2024/017690. P.~K. and H.~L. are supported by the Polish National Science Center (NCN) grant No. 2022/46/E/ST2/00346. P.~K. thanks the EIC theory institute at BNL for its support and hospitality. F.S. is supported by the Laboratory Directed Research and Development of Brookhaven National Laboratory and RIKEN-BNL Research Center, as well as the National Science Foundation (NSF) within the framework of the JETSCAPE collaboration, under grant number OAC-2514008 (CSSI:C-SCAPE). F.S. also acknowledges the Saturated Glue (SURGE) Topical Theory Collaboration, funded by the U.S. Department of Energy, Office of Science, Office of Nuclear Physics. T.S.\ kindly acknowledges the support of the Polish National Science Center (NCN) Grant No.\,2021/43/D/ST2/03375.
\end{acknowledgments}

\bibliographystyle{apsrev4-1}

\bibliography{references.bib}

\begin{thebibliography}{48}%
\makeatletter
\providecommand \@ifxundefined [1]{%
 \@ifx{#1\undefined}
}%
\providecommand \@ifnum [1]{%
 \ifnum #1\expandafter \@firstoftwo
 \else \expandafter \@secondoftwo
 \fi
}%
\providecommand \@ifx [1]{%
 \ifx #1\expandafter \@firstoftwo
 \else \expandafter \@secondoftwo
 \fi
}%
\providecommand \natexlab [1]{#1}%
\providecommand \enquote  [1]{``#1''}%
\providecommand \bibnamefont  [1]{#1}%
\providecommand \bibfnamefont [1]{#1}%
\providecommand \citenamefont [1]{#1}%
\providecommand \href@noop [0]{\@secondoftwo}%
\providecommand \href [0]{\begingroup \@sanitize@url \@href}%
\providecommand \@href[1]{\@@startlink{#1}\@@href}%
\providecommand \@@href[1]{\endgroup#1\@@endlink}%
\providecommand \@sanitize@url [0]{\catcode `\\12\catcode `\$12\catcode
  `\&12\catcode `\#12\catcode `\^12\catcode `\_12\catcode `\%12\relax}%
\providecommand \@@startlink[1]{}%
\providecommand \@@endlink[0]{}%
\providecommand \url  [0]{\begingroup\@sanitize@url \@url }%
\providecommand \@url [1]{\endgroup\@href {#1}{\urlprefix }}%
\providecommand \urlprefix  [0]{URL }%
\providecommand \Eprint [0]{\href }%
\providecommand \doibase [0]{http://dx.doi.org/}%
\providecommand \selectlanguage [0]{\@gobble}%
\providecommand \bibinfo  [0]{\@secondoftwo}%
\providecommand \bibfield  [0]{\@secondoftwo}%
\providecommand \translation [1]{[#1]}%
\providecommand \BibitemOpen [0]{}%
\providecommand \bibitemStop [0]{}%
\providecommand \bibitemNoStop [0]{.\EOS\space}%
\providecommand \EOS [0]{\spacefactor3000\relax}%
\providecommand \BibitemShut  [1]{\csname bibitem#1\endcsname}%
\let\auto@bib@innerbib\@empty
%</preamble>
\bibitem [{\citenamefont {Gribov}\ and\ \citenamefont
  {Lipatov}(1972)}]{Gribov:1972ri}%
  \BibitemOpen
  \bibfield  {author} {\bibinfo {author} {\bibfnamefont {V.~N.}\ \bibnamefont
  {Gribov}}\ and\ \bibinfo {author} {\bibfnamefont {L.~N.}\ \bibnamefont
  {Lipatov}},\ }\href@noop {} {\bibfield  {journal} {\bibinfo  {journal} {Sov.
  J. Nucl. Phys.}\ }\textbf {\bibinfo {volume} {15}},\ \bibinfo {pages} {438}
  (\bibinfo {year} {1972})},\ \bibinfo {note} {[Yad.
  Fiz.15,781(1972)]}\BibitemShut {NoStop}%
%%CITATION = SJNCA,15,438;%%
\bibitem [{\citenamefont {Altarelli}\ and\ \citenamefont
  {Parisi}(1977)}]{Altarelli:1977zs}%
  \BibitemOpen
  \bibfield  {author} {\bibinfo {author} {\bibfnamefont {G.}~\bibnamefont
  {Altarelli}}\ and\ \bibinfo {author} {\bibfnamefont {G.}~\bibnamefont
  {Parisi}},\ }\href {\doibase 10.1016/0550-3213(77)90384-4} {\bibfield
  {journal} {\bibinfo  {journal} {Nucl. Phys.}\ }\textbf {\bibinfo {volume}
  {B126}},\ \bibinfo {pages} {298} (\bibinfo {year} {1977})}\BibitemShut
  {NoStop}%
%%CITATION = NUPHA,B126,298;%%
\bibitem [{\citenamefont {Dokshitzer}(1977)}]{Dokshitzer:1977sg}%
  \BibitemOpen
  \bibfield  {author} {\bibinfo {author} {\bibfnamefont {Y.~L.}\ \bibnamefont
  {Dokshitzer}},\ }\href@noop {} {\bibfield  {journal} {\bibinfo  {journal}
  {Sov. Phys. JETP}\ }\textbf {\bibinfo {volume} {46}},\ \bibinfo {pages} {641}
  (\bibinfo {year} {1977})},\ \bibinfo {note} {[Zh. Eksp. Teor.
  Fiz.73,1216(1977)]}\BibitemShut {NoStop}%
%%CITATION = SPHJA,46,641;%%
\bibitem [{\citenamefont {Lipatov}(1976)}]{Lipatov:1976zz}%
  \BibitemOpen
  \bibfield  {author} {\bibinfo {author} {\bibfnamefont {L.~N.}\ \bibnamefont
  {Lipatov}},\ }\href@noop {} {\bibfield  {journal} {\bibinfo  {journal} {Sov.
  J. Nucl. Phys.}\ }\textbf {\bibinfo {volume} {23}},\ \bibinfo {pages} {338}
  (\bibinfo {year} {1976})}\BibitemShut {NoStop}%
\bibitem [{\citenamefont {Kuraev}\ \emph {et~al.}(1977)\citenamefont {Kuraev},
  \citenamefont {Lipatov},\ and\ \citenamefont {Fadin}}]{Kuraev:1977fs}%
  \BibitemOpen
  \bibfield  {author} {\bibinfo {author} {\bibfnamefont {E.~A.}\ \bibnamefont
  {Kuraev}}, \bibinfo {author} {\bibfnamefont {L.~N.}\ \bibnamefont {Lipatov}},
  \ and\ \bibinfo {author} {\bibfnamefont {V.~S.}\ \bibnamefont {Fadin}},\
  }\href@noop {} {\bibfield  {journal} {\bibinfo  {journal} {Sov. Phys. JETP}\
  }\textbf {\bibinfo {volume} {45}},\ \bibinfo {pages} {199} (\bibinfo {year}
  {1977})}\BibitemShut {NoStop}%
\bibitem [{\citenamefont {Balitsky}\ and\ \citenamefont
  {Lipatov}(1978)}]{Balitsky:1978ic}%
  \BibitemOpen
  \bibfield  {author} {\bibinfo {author} {\bibfnamefont {I.~I.}\ \bibnamefont
  {Balitsky}}\ and\ \bibinfo {author} {\bibfnamefont {L.~N.}\ \bibnamefont
  {Lipatov}},\ }\href@noop {} {\bibfield  {journal} {\bibinfo  {journal} {Sov.
  J. Nucl. Phys.}\ }\textbf {\bibinfo {volume} {28}},\ \bibinfo {pages} {822}
  (\bibinfo {year} {1978})}\BibitemShut {NoStop}%
\bibitem [{\citenamefont {Gribov}\ \emph {et~al.}(1983)\citenamefont {Gribov},
  \citenamefont {Levin},\ and\ \citenamefont {Ryskin}}]{Gribov:1984tu}%
  \BibitemOpen
  \bibfield  {author} {\bibinfo {author} {\bibfnamefont {L.~V.}\ \bibnamefont
  {Gribov}}, \bibinfo {author} {\bibfnamefont {E.~M.}\ \bibnamefont {Levin}}, \
  and\ \bibinfo {author} {\bibfnamefont {M.~G.}\ \bibnamefont {Ryskin}},\
  }\href {\doibase 10.1016/0370-1573(83)90022-4} {\bibfield  {journal}
  {\bibinfo  {journal} {Phys. Rept.}\ }\textbf {\bibinfo {volume} {100}},\
  \bibinfo {pages} {1} (\bibinfo {year} {1983})}\BibitemShut {NoStop}%
\bibitem [{\citenamefont {Mueller}\ and\ \citenamefont
  {Qiu}(1986)}]{Mueller:1985wy}%
  \BibitemOpen
  \bibfield  {author} {\bibinfo {author} {\bibfnamefont {A.~H.}\ \bibnamefont
  {Mueller}}\ and\ \bibinfo {author} {\bibfnamefont {J.-w.}\ \bibnamefont
  {Qiu}},\ }\href {\doibase 10.1016/0550-3213(86)90164-1} {\bibfield  {journal}
  {\bibinfo  {journal} {Nucl. Phys. B}\ }\textbf {\bibinfo {volume} {268}},\
  \bibinfo {pages} {427} (\bibinfo {year} {1986})}\BibitemShut {NoStop}%
\bibitem [{\citenamefont {McLerran}\ and\ \citenamefont
  {Venugopalan}(1994)}]{McLerran:1993ni}%
  \BibitemOpen
  \bibfield  {author} {\bibinfo {author} {\bibfnamefont {L.~D.}\ \bibnamefont
  {McLerran}}\ and\ \bibinfo {author} {\bibfnamefont {R.}~\bibnamefont
  {Venugopalan}},\ }\href {\doibase 10.1103/PhysRevD.49.2233} {\bibfield
  {journal} {\bibinfo  {journal} {Phys. Rev. D}\ }\textbf {\bibinfo {volume}
  {49}},\ \bibinfo {pages} {2233} (\bibinfo {year} {1994})}\BibitemShut
  {NoStop}%
\bibitem [{\citenamefont {Morreale}\ and\ \citenamefont
  {Salazar}(2021)}]{Morreale:2021pnn}%
  \BibitemOpen
  \bibfield  {author} {\bibinfo {author} {\bibfnamefont {A.}~\bibnamefont
  {Morreale}}\ and\ \bibinfo {author} {\bibfnamefont {F.}~\bibnamefont
  {Salazar}},\ }\href {\doibase 10.3390/universe7080312} {\bibfield  {journal}
  {\bibinfo  {journal} {Universe}\ }\textbf {\bibinfo {volume} {7}},\ \bibinfo
  {pages} {312} (\bibinfo {year} {2021})}\BibitemShut {NoStop}%
\bibitem [{\citenamefont {Abdul~Khalek}\ \emph {et~al.}(2022)\citenamefont
  {Abdul~Khalek} \emph {et~al.}}]{AbdulKhalek:2021gbh}%
  \BibitemOpen
  \bibfield  {author} {\bibinfo {author} {\bibfnamefont {R.}~\bibnamefont
  {Abdul~Khalek}} \emph {et~al.},\ }\href {\doibase
  10.1016/j.nuclphysa.2022.122447} {\bibfield  {journal} {\bibinfo  {journal}
  {Nucl. Phys. A}\ }\textbf {\bibinfo {volume} {1026}},\ \bibinfo {pages}
  {122447} (\bibinfo {year} {2022})}\BibitemShut {NoStop}%
\bibitem [{\citenamefont {Gelis}\ \emph {et~al.}(2010)\citenamefont {Gelis},
  \citenamefont {Iancu}, \citenamefont {Jalilian-Marian},\ and\ \citenamefont
  {Venugopalan}}]{Gelis:2010nm}%
  \BibitemOpen
  \bibfield  {author} {\bibinfo {author} {\bibfnamefont {F.}~\bibnamefont
  {Gelis}}, \bibinfo {author} {\bibfnamefont {E.}~\bibnamefont {Iancu}},
  \bibinfo {author} {\bibfnamefont {J.}~\bibnamefont {Jalilian-Marian}}, \ and\
  \bibinfo {author} {\bibfnamefont {R.}~\bibnamefont {Venugopalan}},\ }\href
  {\doibase 10.1146/annurev.nucl.010909.083629} {\bibfield  {journal} {\bibinfo
   {journal} {Ann. Rev. Nucl. Part. Sci.}\ }\textbf {\bibinfo {volume} {60}},\
  \bibinfo {pages} {463} (\bibinfo {year} {2010})}\BibitemShut {NoStop}%
\bibitem [{\citenamefont {Balitsky}(1996)}]{Balitsky:1995ub}%
  \BibitemOpen
  \bibfield  {author} {\bibinfo {author} {\bibfnamefont {I.}~\bibnamefont
  {Balitsky}},\ }\href {\doibase 10.1016/0550-3213(95)00638-9} {\bibfield
  {journal} {\bibinfo  {journal} {Nucl. Phys. B}\ }\textbf {\bibinfo {volume}
  {463}},\ \bibinfo {pages} {99} (\bibinfo {year} {1996})}\BibitemShut
  {NoStop}%
\bibitem [{\citenamefont {Kovchegov}(1999)}]{Kovchegov:1999yj}%
  \BibitemOpen
  \bibfield  {author} {\bibinfo {author} {\bibfnamefont {Y.~V.}\ \bibnamefont
  {Kovchegov}},\ }\href {\doibase 10.1103/PhysRevD.60.034008} {\bibfield
  {journal} {\bibinfo  {journal} {Phys. Rev. D}\ }\textbf {\bibinfo {volume}
  {60}},\ \bibinfo {pages} {034008} (\bibinfo {year} {1999})}\BibitemShut
  {NoStop}%
\bibitem [{\citenamefont {Jalilian-Marian}\ \emph {et~al.}(1997)\citenamefont
  {Jalilian-Marian}, \citenamefont {Kovner}, \citenamefont {Leonidov},\ and\
  \citenamefont {Weigert}}]{JalilianMarian:1997jx}%
  \BibitemOpen
  \bibfield  {author} {\bibinfo {author} {\bibfnamefont {J.}~\bibnamefont
  {Jalilian-Marian}}, \bibinfo {author} {\bibfnamefont {A.}~\bibnamefont
  {Kovner}}, \bibinfo {author} {\bibfnamefont {A.}~\bibnamefont {Leonidov}}, \
  and\ \bibinfo {author} {\bibfnamefont {H.}~\bibnamefont {Weigert}},\ }\href
  {\doibase 10.1016/S0550-3213(97)00440-9} {\bibfield  {journal} {\bibinfo
  {journal} {Nucl. Phys. B}\ }\textbf {\bibinfo {volume} {504}},\ \bibinfo
  {pages} {415} (\bibinfo {year} {1997})}\BibitemShut {NoStop}%
\bibitem [{\citenamefont {Jalilian-Marian}\ \emph {et~al.}(1998)\citenamefont
  {Jalilian-Marian}, \citenamefont {Kovner}, \citenamefont {Leonidov},\ and\
  \citenamefont {Weigert}}]{JalilianMarian:1997gr}%
  \BibitemOpen
  \bibfield  {author} {\bibinfo {author} {\bibfnamefont {J.}~\bibnamefont
  {Jalilian-Marian}}, \bibinfo {author} {\bibfnamefont {A.}~\bibnamefont
  {Kovner}}, \bibinfo {author} {\bibfnamefont {A.}~\bibnamefont {Leonidov}}, \
  and\ \bibinfo {author} {\bibfnamefont {H.}~\bibnamefont {Weigert}},\ }\href
  {\doibase 10.1103/PhysRevD.59.014014} {\bibfield  {journal} {\bibinfo
  {journal} {Phys. Rev. D}\ }\textbf {\bibinfo {volume} {59}},\ \bibinfo
  {pages} {014014} (\bibinfo {year} {1998})}\BibitemShut {NoStop}%
\bibitem [{\citenamefont {Kovner}\ \emph {et~al.}(2000)\citenamefont {Kovner},
  \citenamefont {Milhano},\ and\ \citenamefont {Weigert}}]{Kovner:2000pt}%
  \BibitemOpen
  \bibfield  {author} {\bibinfo {author} {\bibfnamefont {A.}~\bibnamefont
  {Kovner}}, \bibinfo {author} {\bibfnamefont {J.~G.}\ \bibnamefont {Milhano}},
  \ and\ \bibinfo {author} {\bibfnamefont {H.}~\bibnamefont {Weigert}},\ }\href
  {\doibase 10.1103/PhysRevD.62.114005} {\bibfield  {journal} {\bibinfo
  {journal} {Phys. Rev. D}\ }\textbf {\bibinfo {volume} {62}},\ \bibinfo
  {pages} {114005} (\bibinfo {year} {2000})}\BibitemShut {NoStop}%
\bibitem [{\citenamefont {Iancu}\ \emph
  {et~al.}(2001{\natexlab{a}})\citenamefont {Iancu}, \citenamefont {Leonidov},\
  and\ \citenamefont {McLerran}}]{Iancu:2000hn}%
  \BibitemOpen
  \bibfield  {author} {\bibinfo {author} {\bibfnamefont {E.}~\bibnamefont
  {Iancu}}, \bibinfo {author} {\bibfnamefont {A.}~\bibnamefont {Leonidov}}, \
  and\ \bibinfo {author} {\bibfnamefont {L.~D.}\ \bibnamefont {McLerran}},\
  }\href {\doibase 10.1016/S0375-9474(01)00642-X} {\bibfield  {journal}
  {\bibinfo  {journal} {Nucl. Phys. A}\ }\textbf {\bibinfo {volume} {692}},\
  \bibinfo {pages} {583} (\bibinfo {year} {2001}{\natexlab{a}})}\BibitemShut
  {NoStop}%
\bibitem [{\citenamefont {Iancu}\ \emph
  {et~al.}(2001{\natexlab{b}})\citenamefont {Iancu}, \citenamefont {Leonidov},\
  and\ \citenamefont {McLerran}}]{Iancu:2001ad}%
  \BibitemOpen
  \bibfield  {author} {\bibinfo {author} {\bibfnamefont {E.}~\bibnamefont
  {Iancu}}, \bibinfo {author} {\bibfnamefont {A.}~\bibnamefont {Leonidov}}, \
  and\ \bibinfo {author} {\bibfnamefont {L.~D.}\ \bibnamefont {McLerran}},\
  }\href {\doibase 10.1016/S0370-2693(01)00524-X} {\bibfield  {journal}
  {\bibinfo  {journal} {Phys. Lett. B}\ }\textbf {\bibinfo {volume} {510}},\
  \bibinfo {pages} {133} (\bibinfo {year} {2001}{\natexlab{b}})}\BibitemShut
  {NoStop}%
\bibitem [{\citenamefont {Ferreiro}\ \emph {et~al.}(2002)\citenamefont
  {Ferreiro}, \citenamefont {Iancu}, \citenamefont {Leonidov},\ and\
  \citenamefont {McLerran}}]{Ferreiro:2001qy}%
  \BibitemOpen
  \bibfield  {author} {\bibinfo {author} {\bibfnamefont {E.}~\bibnamefont
  {Ferreiro}}, \bibinfo {author} {\bibfnamefont {E.}~\bibnamefont {Iancu}},
  \bibinfo {author} {\bibfnamefont {A.}~\bibnamefont {Leonidov}}, \ and\
  \bibinfo {author} {\bibfnamefont {L.}~\bibnamefont {McLerran}},\ }\href
  {\doibase 10.1016/S0375-9474(01)01329-X} {\bibfield  {journal} {\bibinfo
  {journal} {Nucl. Phys. A}\ }\textbf {\bibinfo {volume} {703}},\ \bibinfo
  {pages} {489} (\bibinfo {year} {2002})}\BibitemShut {NoStop}%
\bibitem [{\citenamefont {Albacete}\ and\ \citenamefont
  {Marquet}(2014)}]{Albacete:2014fwa}%
  \BibitemOpen
  \bibfield  {author} {\bibinfo {author} {\bibfnamefont {J.~L.}\ \bibnamefont
  {Albacete}}\ and\ \bibinfo {author} {\bibfnamefont {C.}~\bibnamefont
  {Marquet}},\ }\href {\doibase 10.1016/j.ppnp.2014.01.004} {\bibfield
  {journal} {\bibinfo  {journal} {Prog. Part. Nucl. Phys.}\ }\textbf {\bibinfo
  {volume} {76}},\ \bibinfo {pages} {1} (\bibinfo {year} {2014})}\BibitemShut
  {NoStop}%
\bibitem [{\citenamefont {Korcyl}\ \emph {et~al.}(2026)\citenamefont {Korcyl},
  \citenamefont {Le}, \citenamefont {Salazar},\ and\ \citenamefont
  {Stebel}}]{Korcyl:2026nrz}%
  \BibitemOpen
  \bibfield  {author} {\bibinfo {author} {\bibfnamefont {P.}~\bibnamefont
  {Korcyl}}, \bibinfo {author} {\bibfnamefont {T.~M.~H.}\ \bibnamefont {Le}},
  \bibinfo {author} {\bibfnamefont {F.}~\bibnamefont {Salazar}}, \ and\
  \bibinfo {author} {\bibfnamefont {T.}~\bibnamefont {Stebel}},\ }\href@noop {}
  {\  (\bibinfo {year} {2026})},\ \Eprint {http://arxiv.org/abs/2607.23485}
  {arXiv:2607.23485 [hep-ph]} \BibitemShut {NoStop}%
\bibitem [{\citenamefont {Shi}\ \emph {et~al.}(2022)\citenamefont {Shi},
  \citenamefont {Wang}, \citenamefont {Wei},\ and\ \citenamefont
  {Xiao}}]{Shi:2021hwx}%
  \BibitemOpen
  \bibfield  {author} {\bibinfo {author} {\bibfnamefont {Y.}~\bibnamefont
  {Shi}}, \bibinfo {author} {\bibfnamefont {L.}~\bibnamefont {Wang}}, \bibinfo
  {author} {\bibfnamefont {S.-Y.}\ \bibnamefont {Wei}}, \ and\ \bibinfo
  {author} {\bibfnamefont {B.-W.}\ \bibnamefont {Xiao}},\ }\href {\doibase
  10.1103/PhysRevLett.128.202302} {\bibfield  {journal} {\bibinfo  {journal}
  {Phys. Rev. Lett.}\ }\textbf {\bibinfo {volume} {128}},\ \bibinfo {pages}
  {202302} (\bibinfo {year} {2022})}\BibitemShut {NoStop}%
\bibitem [{\citenamefont {Blaizot}\ \emph
  {et~al.}(2004{\natexlab{a}})\citenamefont {Blaizot}, \citenamefont {Gelis},\
  and\ \citenamefont {Venugopalan}}]{Blaizot:2004wu}%
  \BibitemOpen
  \bibfield  {author} {\bibinfo {author} {\bibfnamefont {J.~P.}\ \bibnamefont
  {Blaizot}}, \bibinfo {author} {\bibfnamefont {F.}~\bibnamefont {Gelis}}, \
  and\ \bibinfo {author} {\bibfnamefont {R.}~\bibnamefont {Venugopalan}},\
  }\href {\doibase 10.1016/j.nuclphysa.2004.07.005} {\bibfield  {journal}
  {\bibinfo  {journal} {Nucl. Phys. A}\ }\textbf {\bibinfo {volume} {743}},\
  \bibinfo {pages} {13} (\bibinfo {year} {2004}{\natexlab{a}})}\BibitemShut
  {NoStop}%
\bibitem [{\citenamefont {Blaizot}\ \emph
  {et~al.}(2004{\natexlab{b}})\citenamefont {Blaizot}, \citenamefont {Gelis},\
  and\ \citenamefont {Venugopalan}}]{Blaizot:2004wv}%
  \BibitemOpen
  \bibfield  {author} {\bibinfo {author} {\bibfnamefont {J.~P.}\ \bibnamefont
  {Blaizot}}, \bibinfo {author} {\bibfnamefont {F.}~\bibnamefont {Gelis}}, \
  and\ \bibinfo {author} {\bibfnamefont {R.}~\bibnamefont {Venugopalan}},\
  }\href {\doibase 10.1016/j.nuclphysa.2004.07.006} {\bibfield  {journal}
  {\bibinfo  {journal} {Nucl. Phys. A}\ }\textbf {\bibinfo {volume} {743}},\
  \bibinfo {pages} {57} (\bibinfo {year} {2004}{\natexlab{b}})}\BibitemShut
  {NoStop}%
\bibitem [{\citenamefont {Dumitru}\ \emph {et~al.}(2006)\citenamefont
  {Dumitru}, \citenamefont {Hayashigaki},\ and\ \citenamefont
  {Jalilian-Marian}}]{Dumitru:2005gt}%
  \BibitemOpen
  \bibfield  {author} {\bibinfo {author} {\bibfnamefont {A.}~\bibnamefont
  {Dumitru}}, \bibinfo {author} {\bibfnamefont {A.}~\bibnamefont
  {Hayashigaki}}, \ and\ \bibinfo {author} {\bibfnamefont {J.}~\bibnamefont
  {Jalilian-Marian}},\ }\href {\doibase 10.1016/j.nuclphysa.2005.11.014}
  {\bibfield  {journal} {\bibinfo  {journal} {Nucl. Phys. A}\ }\textbf
  {\bibinfo {volume} {765}},\ \bibinfo {pages} {464} (\bibinfo {year}
  {2006})}\BibitemShut {NoStop}%
\bibitem [{\citenamefont {Chirilli}\ \emph {et~al.}(2012)\citenamefont
  {Chirilli}, \citenamefont {Xiao},\ and\ \citenamefont
  {Yuan}}]{Chirilli:2012jd}%
  \BibitemOpen
  \bibfield  {author} {\bibinfo {author} {\bibfnamefont {G.~A.}\ \bibnamefont
  {Chirilli}}, \bibinfo {author} {\bibfnamefont {B.-W.}\ \bibnamefont {Xiao}},
  \ and\ \bibinfo {author} {\bibfnamefont {F.}~\bibnamefont {Yuan}},\ }\href
  {\doibase 10.1103/PhysRevD.86.054005} {\bibfield  {journal} {\bibinfo
  {journal} {Phys. Rev. D}\ }\textbf {\bibinfo {volume} {86}},\ \bibinfo
  {pages} {054005} (\bibinfo {year} {2012})}\BibitemShut {NoStop}%
\bibitem [{\citenamefont {Fujii}\ \emph {et~al.}(2026)\citenamefont {Fujii},
  \citenamefont {Hirano}, \citenamefont {Itakura}, \citenamefont {Nara},\ and\
  \citenamefont {Zhao}}]{Fujii:2026ccu}%
  \BibitemOpen
  \bibfield  {author} {\bibinfo {author} {\bibfnamefont {H.}~\bibnamefont
  {Fujii}}, \bibinfo {author} {\bibfnamefont {T.}~\bibnamefont {Hirano}},
  \bibinfo {author} {\bibfnamefont {K.}~\bibnamefont {Itakura}}, \bibinfo
  {author} {\bibfnamefont {Y.}~\bibnamefont {Nara}}, \ and\ \bibinfo {author}
  {\bibfnamefont {S.}~\bibnamefont {Zhao}},\ }\href@noop {} {\  (\bibinfo
  {year} {2026})},\ \Eprint {http://arxiv.org/abs/2605.15494} {arXiv:2605.15494
  [hep-ph]} \BibitemShut {NoStop}%
\bibitem [{\citenamefont {McLerran}\ and\ \citenamefont
  {Venugopalan}(1998)}]{McLerran:1997fk}%
  \BibitemOpen
  \bibfield  {author} {\bibinfo {author} {\bibfnamefont {L.~D.}\ \bibnamefont
  {McLerran}}\ and\ \bibinfo {author} {\bibfnamefont {R.}~\bibnamefont
  {Venugopalan}},\ }\href {\doibase 10.1016/S0370-2693(98)00214-7} {\bibfield
  {journal} {\bibinfo  {journal} {Phys. Lett. B}\ }\textbf {\bibinfo {volume}
  {424}},\ \bibinfo {pages} {15} (\bibinfo {year} {1998})}\BibitemShut
  {NoStop}%
\bibitem [{\citenamefont {Albacete}\ \emph {et~al.}(2011)\citenamefont
  {Albacete}, \citenamefont {Armesto}, \citenamefont {Milhano}, \citenamefont
  {Quiroga-Arias},\ and\ \citenamefont {Salgado}}]{Albacete:2010sy}%
  \BibitemOpen
  \bibfield  {author} {\bibinfo {author} {\bibfnamefont {J.~L.}\ \bibnamefont
  {Albacete}}, \bibinfo {author} {\bibfnamefont {N.}~\bibnamefont {Armesto}},
  \bibinfo {author} {\bibfnamefont {J.~G.}\ \bibnamefont {Milhano}}, \bibinfo
  {author} {\bibfnamefont {P.}~\bibnamefont {Quiroga-Arias}}, \ and\ \bibinfo
  {author} {\bibfnamefont {C.~A.}\ \bibnamefont {Salgado}},\ }\href {\doibase
  10.1140/epjc/s10052-011-1705-3} {\bibfield  {journal} {\bibinfo  {journal}
  {Eur. Phys. J. C}\ }\textbf {\bibinfo {volume} {71}},\ \bibinfo {pages}
  {1705} (\bibinfo {year} {2011})}\BibitemShut {NoStop}%
\bibitem [{\citenamefont {Lappi}\ and\ \citenamefont
  {M{\"a}ntysaari}(2013)}]{Lappi:2013zma}%
  \BibitemOpen
  \bibfield  {author} {\bibinfo {author} {\bibfnamefont {T.}~\bibnamefont
  {Lappi}}\ and\ \bibinfo {author} {\bibfnamefont {H.}~\bibnamefont
  {M{\"a}ntysaari}},\ }\href {\doibase 10.1103/PhysRevD.88.114020} {\bibfield
  {journal} {\bibinfo  {journal} {Phys. Rev. D}\ }\textbf {\bibinfo {volume}
  {88}},\ \bibinfo {pages} {114020} (\bibinfo {year} {2013})}\BibitemShut
  {NoStop}%
\bibitem [{\citenamefont {Balitsky}(2007)}]{Balitsky:2006wa}%
  \BibitemOpen
  \bibfield  {author} {\bibinfo {author} {\bibfnamefont {I.}~\bibnamefont
  {Balitsky}},\ }\href {\doibase 10.1103/PhysRevD.75.014001} {\bibfield
  {journal} {\bibinfo  {journal} {Phys. Rev. D}\ }\textbf {\bibinfo {volume}
  {75}},\ \bibinfo {pages} {014001} (\bibinfo {year} {2007})}\BibitemShut
  {NoStop}%
\bibitem [{\citenamefont {Motyka}\ and\ \citenamefont
  {Stasto}(2009)}]{Motyka:2009gi}%
  \BibitemOpen
  \bibfield  {author} {\bibinfo {author} {\bibfnamefont {L.}~\bibnamefont
  {Motyka}}\ and\ \bibinfo {author} {\bibfnamefont {A.~M.}\ \bibnamefont
  {Stasto}},\ }\href {\doibase 10.1103/PhysRevD.79.085016} {\bibfield
  {journal} {\bibinfo  {journal} {Phys. Rev. D}\ }\textbf {\bibinfo {volume}
  {79}},\ \bibinfo {pages} {085016} (\bibinfo {year} {2009})}\BibitemShut
  {NoStop}%
\bibitem [{\citenamefont {Beuf}(2014)}]{Beuf:2014uia}%
  \BibitemOpen
  \bibfield  {author} {\bibinfo {author} {\bibfnamefont {G.}~\bibnamefont
  {Beuf}},\ }\href {\doibase 10.1103/PhysRevD.89.074039} {\bibfield  {journal}
  {\bibinfo  {journal} {Phys. Rev. D}\ }\textbf {\bibinfo {volume} {89}},\
  \bibinfo {pages} {074039} (\bibinfo {year} {2014})}\BibitemShut {NoStop}%
\bibitem [{\citenamefont {Beuf}\ \emph {et~al.}(2020)\citenamefont {Beuf},
  \citenamefont {H{\"a}nninen}, \citenamefont {Lappi},\ and\ \citenamefont
  {M{\"a}ntysaari}}]{Beuf:2020dxl}%
  \BibitemOpen
  \bibfield  {author} {\bibinfo {author} {\bibfnamefont {G.}~\bibnamefont
  {Beuf}}, \bibinfo {author} {\bibfnamefont {H.}~\bibnamefont {H{\"a}nninen}},
  \bibinfo {author} {\bibfnamefont {T.}~\bibnamefont {Lappi}}, \ and\ \bibinfo
  {author} {\bibfnamefont {H.}~\bibnamefont {M{\"a}ntysaari}},\ }\href
  {\doibase 10.1103/PhysRevD.102.074028} {\bibfield  {journal} {\bibinfo
  {journal} {Phys. Rev. D}\ }\textbf {\bibinfo {volume} {102}},\ \bibinfo
  {pages} {074028} (\bibinfo {year} {2020})}\BibitemShut {NoStop}%
\bibitem [{\citenamefont {Pumplin}\ \emph {et~al.}(2002)\citenamefont
  {Pumplin}, \citenamefont {Stump}, \citenamefont {Huston}, \citenamefont
  {Lai}, \citenamefont {Nadolsky},\ and\ \citenamefont {Tung}}]{Pumplin_2002}%
  \BibitemOpen
  \bibfield  {author} {\bibinfo {author} {\bibfnamefont {J.}~\bibnamefont
  {Pumplin}}, \bibinfo {author} {\bibfnamefont {D.~R.}\ \bibnamefont {Stump}},
  \bibinfo {author} {\bibfnamefont {J.}~\bibnamefont {Huston}}, \bibinfo
  {author} {\bibfnamefont {H.-L.}\ \bibnamefont {Lai}}, \bibinfo {author}
  {\bibfnamefont {P.}~\bibnamefont {Nadolsky}}, \ and\ \bibinfo {author}
  {\bibfnamefont {W.-K.}\ \bibnamefont {Tung}},\ }\href {\doibase
  10.1088/1126-6708/2002/07/012} {\bibfield  {journal} {\bibinfo  {journal}
  {JHEP}\ }\textbf {\bibinfo {volume} {2002}},\ \bibinfo {pages} {012–012}
  (\bibinfo {year} {2002})}\BibitemShut {NoStop}%
\bibitem [{\citenamefont {Bertone}\ \emph {et~al.}(2017)\citenamefont
  {Bertone}, \citenamefont {Carrazza}, \citenamefont {Hartland}, \citenamefont
  {Nocera},\ and\ \citenamefont {Rojo}}]{Bertone:2017tyb}%
  \BibitemOpen
  \bibfield  {author} {\bibinfo {author} {\bibfnamefont {V.}~\bibnamefont
  {Bertone}}, \bibinfo {author} {\bibfnamefont {S.}~\bibnamefont {Carrazza}},
  \bibinfo {author} {\bibfnamefont {N.~P.}\ \bibnamefont {Hartland}}, \bibinfo
  {author} {\bibfnamefont {E.~R.}\ \bibnamefont {Nocera}}, \ and\ \bibinfo
  {author} {\bibfnamefont {J.}~\bibnamefont {Rojo}} (\bibinfo {collaboration}
  {NNPDF}),\ }\href {\doibase 10.1140/epjc/s10052-017-5088-y} {\bibfield
  {journal} {\bibinfo  {journal} {Eur. Phys. J. C}\ }\textbf {\bibinfo {volume}
  {77}},\ \bibinfo {pages} {516} (\bibinfo {year} {2017})}\BibitemShut
  {NoStop}%
\bibitem [{\citenamefont {Golec-Biernat}\ and\ \citenamefont
  {Wusthoff}(1998)}]{Golec-Biernat:1998zce}%
  \BibitemOpen
  \bibfield  {author} {\bibinfo {author} {\bibfnamefont {K.~J.}\ \bibnamefont
  {Golec-Biernat}}\ and\ \bibinfo {author} {\bibfnamefont {M.}~\bibnamefont
  {Wusthoff}},\ }\href {\doibase 10.1103/PhysRevD.59.014017} {\bibfield
  {journal} {\bibinfo  {journal} {Phys. Rev. D}\ }\textbf {\bibinfo {volume}
  {59}},\ \bibinfo {pages} {014017} (\bibinfo {year} {1998})}\BibitemShut
  {NoStop}%
\bibitem [{\citenamefont {Abramowicz}\ \emph {et~al.}(2015)\citenamefont
  {Abramowicz} \emph {et~al.}}]{H1:2015ubc}%
  \BibitemOpen
  \bibfield  {author} {\bibinfo {author} {\bibfnamefont {H.}~\bibnamefont
  {Abramowicz}} \emph {et~al.} (\bibinfo {collaboration} {H1, ZEUS}),\ }\href
  {\doibase 10.1140/epjc/s10052-015-3710-4} {\bibfield  {journal} {\bibinfo
  {journal} {Eur. Phys. J. C}\ }\textbf {\bibinfo {volume} {75}},\ \bibinfo
  {pages} {580} (\bibinfo {year} {2015})}\BibitemShut {NoStop}%
\bibitem [{\citenamefont {Arsene}\ \emph {et~al.}(2004)\citenamefont {Arsene}
  \emph {et~al.}}]{BRAHMS:2004xry}%
  \BibitemOpen
  \bibfield  {author} {\bibinfo {author} {\bibfnamefont {I.}~\bibnamefont
  {Arsene}} \emph {et~al.} (\bibinfo {collaboration} {BRAHMS}),\ }\href
  {\doibase 10.1103/PhysRevLett.93.242303} {\bibfield  {journal} {\bibinfo
  {journal} {Phys. Rev. Lett.}\ }\textbf {\bibinfo {volume} {93}},\ \bibinfo
  {pages} {242303} (\bibinfo {year} {2004})}\BibitemShut {NoStop}%
\bibitem [{\citenamefont {Adams}\ \emph {et~al.}(2006)\citenamefont {Adams}
  \emph {et~al.}}]{STAR:2006dgg}%
  \BibitemOpen
  \bibfield  {author} {\bibinfo {author} {\bibfnamefont {J.}~\bibnamefont
  {Adams}} \emph {et~al.} (\bibinfo {collaboration} {STAR}),\ }\href {\doibase
  10.1103/PhysRevLett.97.152302} {\bibfield  {journal} {\bibinfo  {journal}
  {Phys. Rev. Lett.}\ }\textbf {\bibinfo {volume} {97}},\ \bibinfo {pages}
  {152302} (\bibinfo {year} {2006})}\BibitemShut {NoStop}%
\bibitem [{\citenamefont {Aaij}\ \emph
  {et~al.}(2022{\natexlab{a}})\citenamefont {Aaij} \emph
  {et~al.}}]{LHCb:2021vww}%
  \BibitemOpen
  \bibfield  {author} {\bibinfo {author} {\bibfnamefont {R.}~\bibnamefont
  {Aaij}} \emph {et~al.} (\bibinfo {collaboration} {LHCb}),\ }\href {\doibase
  10.1103/PhysRevLett.128.142004} {\bibfield  {journal} {\bibinfo  {journal}
  {Phys. Rev. Lett.}\ }\textbf {\bibinfo {volume} {128}},\ \bibinfo {pages}
  {142004} (\bibinfo {year} {2022}{\natexlab{a}})}\BibitemShut {NoStop}%
\bibitem [{\citenamefont {Aaij}\ \emph
  {et~al.}(2022{\natexlab{b}})\citenamefont {Aaij} \emph
  {et~al.}}]{LHCb:2021abm}%
  \BibitemOpen
  \bibfield  {author} {\bibinfo {author} {\bibfnamefont {R.}~\bibnamefont
  {Aaij}} \emph {et~al.} (\bibinfo {collaboration} {LHCb}),\ }\href {\doibase
  10.1007/JHEP01(2022)166} {\bibfield  {journal} {\bibinfo  {journal} {JHEP}\
  }\textbf {\bibinfo {volume} {01}},\ \bibinfo {pages} {166} (\bibinfo {year}
  {2022}{\natexlab{b}})}\BibitemShut {NoStop}%
\bibitem [{\citenamefont {Duclou{\'e}}\ \emph {et~al.}(2019)\citenamefont
  {Duclou{\'e}}, \citenamefont {Iancu}, \citenamefont {Mueller}, \citenamefont
  {Soyez},\ and\ \citenamefont {Triantafyllopoulos}}]{Ducloue:2019ezk}%
  \BibitemOpen
  \bibfield  {author} {\bibinfo {author} {\bibfnamefont {B.}~\bibnamefont
  {Duclou{\'e}}}, \bibinfo {author} {\bibfnamefont {E.}~\bibnamefont {Iancu}},
  \bibinfo {author} {\bibfnamefont {A.~H.}\ \bibnamefont {Mueller}}, \bibinfo
  {author} {\bibfnamefont {G.}~\bibnamefont {Soyez}}, \ and\ \bibinfo {author}
  {\bibfnamefont {D.~N.}\ \bibnamefont {Triantafyllopoulos}},\ }\href {\doibase
  10.1007/JHEP04(2019)081} {\bibfield  {journal} {\bibinfo  {journal} {JHEP}\
  }\textbf {\bibinfo {volume} {04}},\ \bibinfo {pages} {081} (\bibinfo {year}
  {2019})}\BibitemShut {NoStop}%
\bibitem [{\citenamefont {Casuga}\ \emph {et~al.}(2025)\citenamefont {Casuga},
  \citenamefont {H{\"a}nninen},\ and\ \citenamefont
  {M{\"a}ntysaari}}]{Casuga:2025etc}%
  \BibitemOpen
  \bibfield  {author} {\bibinfo {author} {\bibfnamefont {C.}~\bibnamefont
  {Casuga}}, \bibinfo {author} {\bibfnamefont {H.}~\bibnamefont
  {H{\"a}nninen}}, \ and\ \bibinfo {author} {\bibfnamefont {H.}~\bibnamefont
  {M{\"a}ntysaari}},\ }\href {\doibase 10.1103/54zd-hyvg} {\bibfield  {journal}
  {\bibinfo  {journal} {Phys. Rev. D}\ }\textbf {\bibinfo {volume} {112}},\
  \bibinfo {pages} {034003} (\bibinfo {year} {2025})}\BibitemShut {NoStop}%
\bibitem [{\citenamefont {Pumplin}\ \emph
  {et~al.}(2001{\natexlab{a}})\citenamefont {Pumplin}, \citenamefont {Stump},\
  and\ \citenamefont {Tung}}]{Pumplin:2000vx}%
  \BibitemOpen
  \bibfield  {author} {\bibinfo {author} {\bibfnamefont {J.}~\bibnamefont
  {Pumplin}}, \bibinfo {author} {\bibfnamefont {D.~R.}\ \bibnamefont {Stump}},
  \ and\ \bibinfo {author} {\bibfnamefont {W.~K.}\ \bibnamefont {Tung}},\
  }\href {\doibase 10.1103/PhysRevD.65.014011} {\bibfield  {journal} {\bibinfo
  {journal} {Phys. Rev. D}\ }\textbf {\bibinfo {volume} {65}},\ \bibinfo
  {pages} {014011} (\bibinfo {year} {2001}{\natexlab{a}})}\BibitemShut
  {NoStop}%
\bibitem [{\citenamefont {Pumplin}\ \emph
  {et~al.}(2001{\natexlab{b}})\citenamefont {Pumplin}, \citenamefont {Stump},
  \citenamefont {Brock}, \citenamefont {Casey}, \citenamefont {Huston},
  \citenamefont {Kalk}, \citenamefont {Lai},\ and\ \citenamefont
  {Tung}}]{Pumplin:2001ct}%
  \BibitemOpen
  \bibfield  {author} {\bibinfo {author} {\bibfnamefont {J.}~\bibnamefont
  {Pumplin}}, \bibinfo {author} {\bibfnamefont {D.}~\bibnamefont {Stump}},
  \bibinfo {author} {\bibfnamefont {R.}~\bibnamefont {Brock}}, \bibinfo
  {author} {\bibfnamefont {D.}~\bibnamefont {Casey}}, \bibinfo {author}
  {\bibfnamefont {J.}~\bibnamefont {Huston}}, \bibinfo {author} {\bibfnamefont
  {J.}~\bibnamefont {Kalk}}, \bibinfo {author} {\bibfnamefont {H.~L.}\
  \bibnamefont {Lai}}, \ and\ \bibinfo {author} {\bibfnamefont {W.~K.}\
  \bibnamefont {Tung}},\ }\href {\doibase 10.1103/PhysRevD.65.014013}
  {\bibfield  {journal} {\bibinfo  {journal} {Phys. Rev. D}\ }\textbf {\bibinfo
  {volume} {65}},\ \bibinfo {pages} {014013} (\bibinfo {year}
  {2001}{\natexlab{b}})}\BibitemShut {NoStop}%
\bibitem [{\citenamefont {Cougoulic}\ \emph {et~al.}(2025)\citenamefont
  {Cougoulic}, \citenamefont {Korcyl},\ and\ \citenamefont
  {Stebel}}]{Cougoulic:2024jnd}%
  \BibitemOpen
  \bibfield  {author} {\bibinfo {author} {\bibfnamefont {F.}~\bibnamefont
  {Cougoulic}}, \bibinfo {author} {\bibfnamefont {P.}~\bibnamefont {Korcyl}}, \
  and\ \bibinfo {author} {\bibfnamefont {T.}~\bibnamefont {Stebel}},\ }\href
  {\doibase 10.1016/j.cpc.2025.109616} {\bibfield  {journal} {\bibinfo
  {journal} {Comput. Phys. Commun.}\ }\textbf {\bibinfo {volume} {313}},\
  \bibinfo {pages} {109616} (\bibinfo {year} {2025})}\BibitemShut {NoStop}%
\end{thebibliography}%

\end{document}